# Ground-based verification and data processing of Yutu rover Active Particle-induced X-ray Spectrometer *

Guo Dong-Ya(郭东亚)[1], Wang Huan-Yu(王焕玉)[1], Peng Wen-Xi(彭文溪)[1;1)], Cui Xing-Zhu(崔兴柱)[1], Zhang Cheng-Mo(张承模)[1], Liu Ya-Qing(刘雅清)[1], Liang Xiao-Hua(梁晓华)[1], Dong Yi-Fan(董亦凡)[1], Wang Jin-Zhou(汪锦州)[1], Gao Min(高旻)[1], Yang Jia-Wei(杨家卫)[1], Zhang Jia-Yu(张家宇)[1], Li Chun-Lai(李春来)[2], Zou Yong-Liao(邹永廖)[2], Zhang Guang-Liang(张广良)[2], Zhang Li-Yan(张力岩)[2], Fu Xiao-Hui(付晓辉)[2]

[1] Key Laboratory of Particle Astrophysics, Institute of High Energy Physics, Chinese Academy of Sciences, Beijing 100049, China

[2] National Astronomical Observatories, Chinese Academy of Sciences, Beijing 100012, China

**Abstract:** The Active Particle-induced X-ray Spectrometer (APXS) is one of the payloads on board the Yutu rover of Chang'E-3 mission. In order to assess the instrumental performance of APXS, a ground verification test was done for two unknown samples (basaltic rock, mixed powder sample). In this paper, the details of the experiment configurations and data analysis method are presented. The results show that the elemental abundance of major elements can be well determined by the APXS with relative deviations < 15 wt. % (detection distance = 30 mm, acquisition time = 30 min). The derived detection limit of each major element is inversely proportional to acquisition time and directly proportional to detection distance, suggesting that the appropriate distance should be < 50mm.

## 1 Introduction

The Active Particle-induced X-ray Spectrometer (APXS) system, on board the Yutu rover of the Chang'E-3 satellite, is designed to determine the element composition and abundance of lunar soils and rocks. It consists of a sensor head, a Radioisotope Heater Unit (RHU) and calibration equipment (as shown in Figure 1). The sensor head is mounted on a robotic arm at the front of the Yutu rover. It has a Silicon Drift Detector (SDD, 7 mm$^2$ effective areas) in the center surrounded by eight excitation sources. The RHU and the calibration equipment are both mounted on the front panel of the rover. The RHU is used to keep the sensor head warm during the moon night, and the calibration equipment, which has a basalt disk, is used for the on-orbit calibration [1].

The working principle of the Chang'E-3 APXS is similar to other X-ray fluorescence spectrometers. Elementary incident X-rays are provided by a combination of four $^{55}$Fe sources (294 mCi in total, emitting X-rays of energy 5.90 keV and 6.49 keV) and four $^{109}$Cd sources (13.4 mCi in total, emitting X-rays of energy 22.1keV, 24.9 keV, 25.5 keV and 88 keV), like those used in the Viking and Beagle-2 missions to Mars [2, 3]. In detection mode, as shown in Figure 2, the robotic arm deploys the sensor head to a target on the lunar surface, and X-rays emitted by the radioactive sources bombard atoms in the lunar soil. These atoms will be excited and produce characteristic X-rays in the de-excitation process. The SDD and electronics system then record and accumulate the characteristic X-rays. From the accumulated characteristic X-ray spectrum, we can infer the element composition and abundance of the lunar soil.

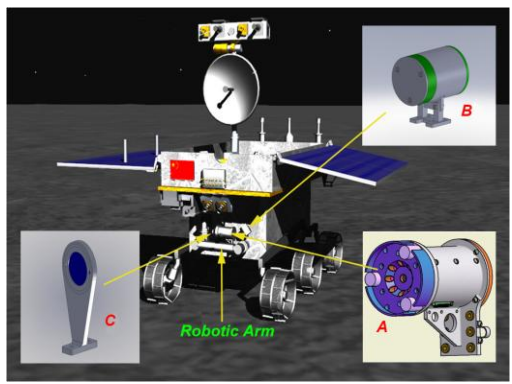

Figure 1. Model of Yutu rover and APXS components (A: sensor head; B: RHU; C: calibration equipment)

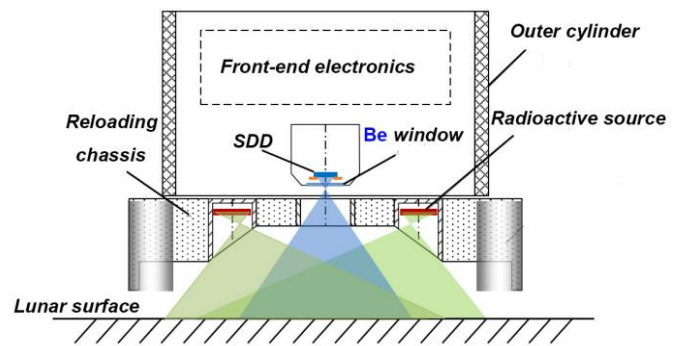

Figure 2. Working principle of APXS

* Supported by National Science and Technology Major Project (Chang'E-3 Active Particle-induced X-ray Spectrometer)
1) E-mail: pengwx@ihep.ac.cn

The main technical specifications of APXS are shown in Table 1.

Table 1. Main technical specifications of APXS

| Technical specification | Description |
| --- | --- |
| Type of detector | Silicon Drift Detector (SDD) |
| Effective area | 7mm$^2$ |
| Energy resolution | <140eV@5.9keV |
| Energy range | 0.5~20keV |
| Detection distance | 10~30mm |
| Active source | 4×70mCi $^{55}$Fe + 4×5mCi $^{109}$Cd |
| Weight(sensor head only) | 754g |

Before being integrated on the rover, a blind test was done for APXS to validate its quantitative analysis ability for the major elements of igneous rocks. In this paper, the experiment setup, data processing method and results are described.

## 2 Verification test

The main goals of the verification test are the following:
1) To assess the accuracy of APXS quantitative analysis;
2) To validate the data processing procedure and methodology;
3) To study the influence of the data acquisition time and detection distance on the limit of detection (LOD) of major elements.

### 2.1 Testing configuration

The test was done in a vacuum chamber. As shown in Figure 3, the sensor head of the APXS flight model (FM) was mounted on a vertical motion worktable, which can change the distance between the sensor head and target (detection distance). The targets (powder or rock) were prepared and placed under the sensor head.

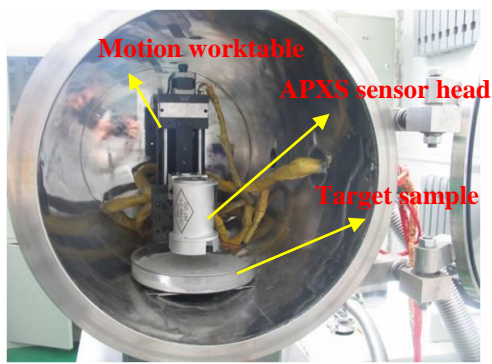

Figure 3. The vacuum chamber containing the sensor head, target and motion worktable

### 2.2 Sample preparation

A basalt rock was collected in Zhangbei, Hebei Province, China. This rock was cut into two parts. One part was crushed into powder and mixed with other mineral powders, labeled 'unknown mixed powder sample'. The other was cut into a 13cm*13cm*2 cm cube, and labeled 'unknown rock sample'. Two samples will allow us to assess the elemental analysis capability of APXS for both lunar rock and powder samples. Both prepared samples were sent to several laboratories for the compositional analysis of major elements. The results are shown in Table 2; the authors of this paper did not know these values in advance.

In addition, a geochemistry standard powder sample, GBW07105 (with certified abundance of major elements as listed in Table 2) was used in this test.

During the test, the powder samples were put into a glass tray and kept with a flat surface of thickness around 10 mm, which is equal to an infinitely thick target for incident X-rays.

Table 2. Verified compositions of test samples

| Sample | | Sample Compositions (wt. %) | | | | | | |
| --- | --- | --- | --- | --- | --- | --- | --- | --- |
| | | *Mg* | *Al* | *Si* | *K* | *Ca* | *Ti* | *Fe* |
| Rock | Mean | 5.50 | 7.15 | 20.58 | 1.70 | 6.56 | 1.55 | 9.57 |
| | σ | 0.04 | 0.02 | 0.03 | 0.02 | 0.03 | 0.01 | 0.04 |
| Powder | Mean | 5.40 | 6.35 | 17.96 | 1.50 | 9.67 | 1.36 | 8.27 |
| | σ | 0.01 | 0.01 | 0.02 | 0.03 | 0.01 | 0.01 | 0.01 |
| GBW 07105 | Mean | 4.72 | 7.37 | 21.01 | 1.94 | 6.34 | 1.43 | 9.44 |
| | σ | 0.16 | 0.11 | 0.07 | 0.07 | 0.10 | 0.04 | 0.13 |

### 2.3 Test procedure

When the pressure of the vacuum chamber was lower than 10 Pa, the following tests were done:
1) For each sample (the two unknown samples and GBW07105), a spectrum of 30 minutes was accumulated at a detection distance of 30mm.
2) With the unknown mixed powder sample as the target, the detection distance was first changed gradually from 10 mm to 110 mm (10 mm, 20 mm, 30 mm, 40 mm, 50 mm, 70 mm, 90 mm and 110 mm), while the data acquisition time was still 30 min. Then, the detection distance was kept at 30 mm, but the acquisition time for each spectrum changed from 10 min to 90 min (10 min, 20 min, 30 min, 45 min, 60 min and 90 min).

## 3 Data processing

### 3.1 Data processing procedure

The data processing steps for the APXS spectra are shown in Figure 4.

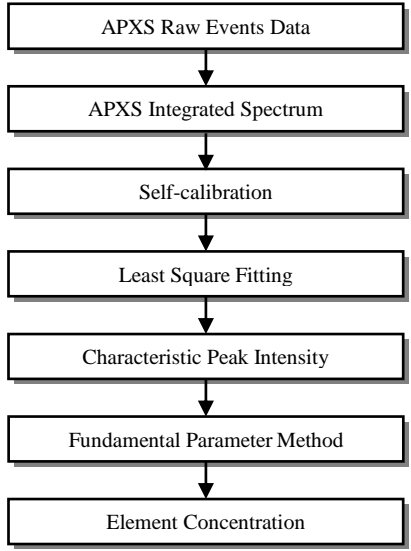

Figure 4. Data processing procedure for APXS RAW data

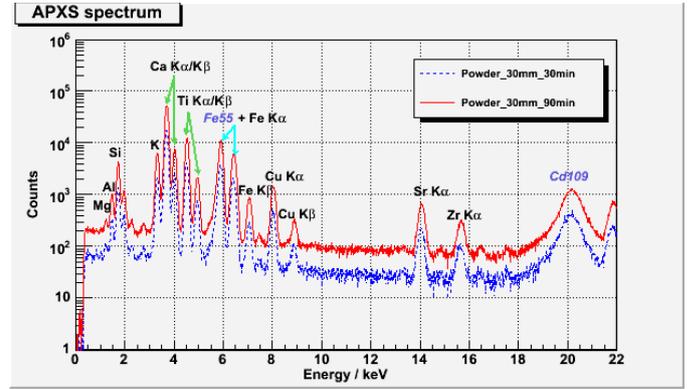

Figure 5. Spectra of unknown mixed powder sample (dashed line: acquisition time is 30 minutes; solid line: acquisition time is 90 minutes)

A spectrum with 2048 ADC channels can be integrated directly from the APXS raw event data. In general, geological samples contain several major elements, so we can do self-calibration as follows. First, according to the stand-alone energy linear pre-calibration of APXS, the characteristic peaks of Ca Kα, Ti Kβ and Fe Kβ could be identified easily; these were then fitted by Gaussian functions to get their positions and FWHMs. The fitting results show that the peak positions and FWHMs of different spectra obtained in the verification test are almost the same for each element (the temperature drift was very small during the test). Therefore the energy calibration and energy resolution calibration can be calculated using their mean values, by equations (1) and (2):

$$E(\text{eV}) = K_1 * Ch + C_1, \quad (1)$$

$$FWHM(\text{eV}) = \sqrt{K_2 * E(\text{eV}) + C_2}, \quad (2)$$

where $K_1$ and $K_2$ are the slope parameters, $C_1$ and $C_2$ are the intercept parameters, $Ch$ is the ADC channel number and $E$ (eV) is the energy in eV. The calibration results in $K_1$=11.639, $K_2$=3.1061, $C_1$=27.48, $C_2$=8992.6, which were used for all the spectra.

Two typical spectra of the unknown mixed powder sample are presented in Figure 5. As shown in that figure, the characteristic peaks of the major elements (Mg, Al, Si, K, Ca, Ti and Fe) in the sample can easily be distinguished in the spectra. Besides, some peaks of the trace elements, such as Sr, Zr and so on, can also be discovered in the spectra. In this study, we just focus on the major elements.

Spectrum fitting is a key step of the data processing procedure.

As shown in Figure 5, the typical spectrum has a continuous background superimposed with some Gaussian peaks, which are characteristic peaks, escape peaks and backscattering peaks.

Theoretically, the background of the APXS spectrum comes from multi-scattering X-rays and secondary electron bremsstrahlung radiation processes, which are related to the sample compositions, radioactive sources and geometric layout. It is difficult to construct an accurate analytical model based on physical theory. In our data processing, instead of studying the physical model of the background, the SNIP (Sensitive Nonlinear Iterative Peak-clipping) algorithm [4] is implemented to estimate the spectral background, as was first proposed by C. G. Ryan [5], and improved by M. Morháč [6].

In general, the response of silicon detectors to mono-energetic X-rays consists of three main components: Gaussian total energy peak, low-energy tail and escape peak, described by the HYPERMET equation [7-9], as shown in Equations (3-5):

Gaussian total energy peak:

$$G(i) = H_G \cdot \frac{1}{\sqrt{2\pi}\sigma} \exp\left[-\frac{(i-i_0)^2}{2\sigma^2}\right], \quad (3)$$

Exponential Tail:

$$E(i) = H_E \exp\left(\frac{i-i_0}{\beta}\right) \cdot \frac{1}{2} \text{erfc}\left(\frac{i-i_0}{\sigma\sqrt{2}} + \frac{\sigma}{\beta\sqrt{2}}\right) \cdot \frac{1}{2} \text{erfc}\left(\frac{i_s-i}{\sigma\sqrt{2}}\right), \quad (4)$$

Escape Peak:

$$G_{\text{Esc}}(i) = H_{\text{Esc}} \cdot \frac{1}{\sqrt{2\pi}\sigma} \exp\left[-\frac{(i-i_{\text{esc}})^2}{2\sigma^2}\right]. \quad (5)$$

$H$ represents the intensity of each component, $i_0$ is the position of the total energy peak, $\sigma$ is the standard deviation of the Gaussian peak, and $\beta$ is a parameter of the exponential tail; $\sigma$ and $\beta$ reflect the wide extent of the spectral shape. $i_s$ is the front cutoff position of the exponential function, $i_{\text{esc}}$ is the position of the escape peak.

A fitting code (CAPXS) based on these models was developed in the framework of ROOT. Fitting functions used for the different element characteristic peaks are shown in Table 3.

Table 3. Fitting functions for different elements

| Function Classes | Fitting Function | Applicable Elements |
|---|---|---|
| Func1 | Gaussian | Mg, Al, Si, Ni, Rb, Sr, Y, Zr, Nb |
| Func2 | Gaussian + Exponent Tail + Escape Peak | K, Ca, Ti, Fe, Cu |
| Func3 | Gaussian + Exponent Tail + Escape Peak | $^{55}$Fe |

As several peaks overlap, some parameters should be provided as constraints, such as the ratio of total energy peak to escape peak, and the ratio of Kα peak to Kβ peak. All these parameters are determined by other ground tests (These tests results will be discussed in future papers). Figure 6 shows the fitting result of the unknown rock sample.

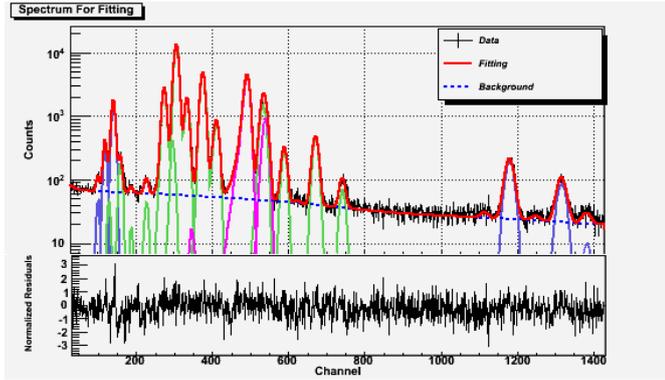

Figure 6. Spectrum fitting and fitting residuals of the unknown rock sample (in the top graph, dashed line: background; solid line: fitting result)

Once we have the characteristic peak areas of each element in both the unknown sample and standard sample, the elemental concentrations of the unknown sample can be derived by the Fundamental Parameter algorithm (FP).

FP is based on the X-ray fluorescence theoretical production model, in which the real geometric model of APXS was constructed. The model correction factor $\varepsilon_i$ is used to characterize the relationship between the measured and theoretical intensity, which is calculated by Equation (6).

$$\varepsilon_i = \frac{I_{i,m,S}}{I_{i,t,S}}, \qquad (6)$$

where $I_{i,m,S}$ and $I_{i,t,S}$ represent the measured and theoretical intensities respectively of element i in the standard sample.

The procedure of the FP algorithm is shown in Figure 6. Firstly, the characteristic peak intensities of the unknown sample are normalized and compared with the standard sample to get a set of initial concentrations $C_i$. Then the relative intensities of the unknown sample are calculated using the initial concentration and theoretical model. According to the model correction factor, the measured and theoretical relative intensities, we can get a set of corrected concentrations $C_i^{'}$. This process is reiterated until the difference between $C_i$ and $C_i^{'}$ is less than 0.0008 (an appropriate and small enough number, such that neither the number of iterations nor the deviation are too big).

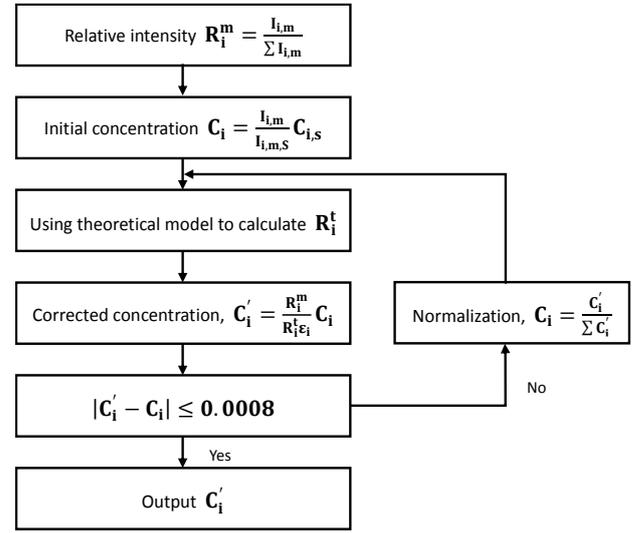

Figure 7. Calculation process of fundamental parameter method [10]

In Figure 7, $I_{i,m}$ is the measured intensity of element $i$ in the unknown sample. $R_i^m$ and $R_i^t$ are the measured and theoretical normalized intensities of element $i$ in the unknown sample separately. $C_{i,S}$ is the weight percentage of element $i$ in the standard sample.

## 3.2 Analysis result

Following the data processing procedure as described above, the quantitative results of the two unknown samples are derived, and then compared with the certified value, as shown in Table 4.

Table 4. Results of quantitative analysis

| element | unknown rock sample | | | unknown mixed powder sample | | |
|---|---|---|---|---|---|---|
| | Derived Value (wt. %) | Certified Value (wt. %) | Relative error (%) | Derived Value (wt. %) | Certified Value (wt. %) | Relative error (%) |
| Mg | 4.80 | 5.50 | 12.7 | 5.45 | 5.40 | 0.93 |
| Al | 8.06 | 7.10 | 13.5 | 6.77 | 6.35 | 6.61 |
| Si | 21.26 | 20.58 | 3.3 | 19.31 | 17.96 | 7.52 |
| K | 1.95 | 1.70 | 14.7 | 1.63 | 1.50 | 8.67 |
| Ca | 6.91 | 6.56 | 5.3 | 9.95 | 9.67 | 2.90 |
| Ti | 1.46 | 1.55 | 5.8 | 1.38 | 1.36 | 1.47 |
| Fe | 8.5 | 9.57 | 11.2 | 9.04 | 8.27 | 9.31 |

The relative deviation is normally used as the criterion of elemental determination accuracy, and is expressed in Equation (7):

$$R_i = \left|\frac{C_{a,i} - C_{c,i}}{C_{c,i}}\right| \times 100\%, \qquad (7)$$

where $C_{a,i}$ is the analytic result and $C_{c,i}$ is the certified value. The main sources of relative error are the statistical error, theoretical model error, sample preparation error and so on. According to the results, for both unknown rock and powder samples, the relative deviation of derived concentration for each major element is less than 15 wt. % at the distance of 30 mm with 30 min data acquisition time. Therefore, the elemental abundances of major elements could be well determined by the APXS with the data processing method described here.

### 3.3 Limit of detection (LOD)

The LODs of APXS for different elements were calculated by using Equations (8, 9) [3, 11]:

$$LOD = \frac{3\sqrt{2}}{S}\sqrt{\frac{I_b + I_{ip}}{t}}, \quad (8)$$

$$S = \frac{I_p}{C_i}, \quad (9)$$

where $I_b$ is the background count rate in channel [$i_0$-3$\sigma$, $i_0$+3$\sigma$], $i_0$ and $\sigma$ are the peak position and standard deviation of the analyzed characteristic line, $I_{ip}$ is the interference peak count rate in this channel range, $I_p$ is the peak count rate and $t$ is the data acquisition time. $S$ is defined as element sensitivity and $C_i$ is the concentration of element $i$.

According to Equation (8), the LOD is related to the element sensitivity, background count rate and data acquisition time.

When the detection distance is fixed at 30mm, the LOD of each element under different acquisition time is shown in Table 5 and Figure 8 (only the fitting error and statistical error were considered).

As showed in Figure 8, the detection capability of APXS for low atomic number elements such as Mg is poorer than for high atomic elements, because of the low excitation cross-section of $^{55}$Fe source for these elements. In addition, the LODs of all the elements are improved with the lengthening of acquisition time.

Table 5. LOD of each element under different detection time

| Acquisition time /min | Mg | | Al | | Si | | K | | Ca | | Ti | | Fe | |
|---|---|---|---|---|---|---|---|---|---|---|---|---|---|---|
| | **LOD** /wt. % | Error /wt. % | **LOD** /wt. % | Error /wt. % | **LOD** /wt. % | Error /ppm | **LOD** /ppm | Error /ppm | **LOD** /ppm | Error /ppm | **LOD** /ppm | Error /ppm | **LOD** /wt. % | Error /ppm |
| 10 | **4.47** | 1.92 | **1.00** | 0.09 | **0.47** | 153.05 | **231.1** | 9.01 | **186.4** | 3.33 | **92** | 2.73 | **0.58** | 354.54 |
| 20 | **2.26** | 0.40 | **0.70** | 0.05 | **0.31** | 69.34 | **162.9** | 4.85 | **127.6** | 1.66 | **63.1** | 1.4 | **0.43** | 183.96 |
| 30 | **2.24** | 0.46 | **0.44** | 0.02 | **0.26** | 47.97 | **132.1** | 3.66 | **98.7** | 1.13 | **52.8** | 1.03 | **0.34** | 121.21 |
| 45 | **1.78** | 0.31 | **0.48** | 0.02 | **0.21** | 32.71 | **108** | 2.83 | **86.6** | 0.79 | **41.5** | 0.74 | **0.26** | 77.69 |
| 60 | **1.61** | 0.27 | **0.41** | 0.02 | **0.18** | 25.33 | **90.1** | 2.22 | **73.1** | 0.60 | **35.8** | 0.59 | **0.23** | 58.75 |
| 90 | **1.25** | 0.17 | **0.34** | 0.01 | **0.15** | 16.97 | **73.2** | 1.71 | **59.6** | 0.42 | **29.7** | 0.44 | **0.19** | 39.71 |

Table 6. LOD of each element under different detection distance

| Detection distance /mm | Mg | | Al | | Si | | K | | Ca | | Ti | | Fe | |
|---|---|---|---|---|---|---|---|---|---|---|---|---|---|---|
| | **LOD** /wt. % | Error /wt. % | **LOD** /wt. % | Error /wt. % | **LOD** /wt. % | Error /ppm | **LOD** /ppm | Error /ppm | **LOD** /ppm | Error /ppm | **LOD** /ppm | Error /ppm | **LOD** /wt. % | Error /ppm |
| 10 | **1.20** | 0.16 | **0.37** | 0.01 | **0.17** | 22.67 | **91.6** | 1.92 | **68.8** | 0.56 | **32.4** | 0.50 | **0.18** | 42.19 |
| 20 | **1.67** | 0.31 | **0.49** | 0.02 | **0.21** | 34.59 | **102.5** | 2.59 | **81.8** | 0.80 | **38.1** | 0.69 | **0.23** | 66.36 |
| 30 | **2.24** | 0.46 | **0.44** | 0.02 | **0.26** | 47.97 | **132.1** | 3.66 | **98.7** | 1.13 | **52.8** | 1.03 | **0.34** | 121.21 |
| 40 | **2.36** | 0.53 | **0.70** | 0.04 | **0.33** | 70.05 | **154.5** | 4.94 | **124.6** | 1.53 | **60.7** | 1.36 | **0.43** | 194.15 |
| 50 | **2.82** | 0.74 | **0.84** | 0.06 | **0.39** | 96.5 | **194.5** | 7.72 | **160.4** | 2.15 | **76** | 2.02 | **0.51** | 247.43 |
| 70 | **4.28** | 1.40 | **1.25** | 0.11 | **0.60** | 172.42 | **28.1** | 9.71 | **214.1** | 3.16 | **112** | 2.78 | **0.89** | 608.33 |
| 90 | **5.01** | 1.91 | **1.33** | 0.13 | **0.84** | 312.58 | **426.9** | 17.04 | **303** | 4.8 | **163.8** | 4.43 | **1.29** | 1225.59 |
| 110 | **5.95** | 2.18 | **1.90** | 0.26 | **1.22** | 553.94 | **546.6** | 28.16 | **390.9** | 7.14 | **224** | 7.49 | **1.95** | 2369.45 |

The LOD is also sensitive to the detection distance. When the data acquisition time is fixed at 30 min, the detection limit of each element for different detection distances is shown in Table 6 and Figure 9.

Figure 9 shows that the LOD of each element increases with an increase in detection distance. When the detection distance is longer than 50 mm, the LOD of Mg is around 3 wt. % (MgO is about 5 wt. %), which is also the lower limit of Mg content in lunar soil. Figure 10 gives the average Mg content of lunar soil returned by the Apollo and Luna missions [12]. Therefore, in order to analyze the major rock-forming elements in a lunar sample, the detection distance should be controlled within 50 mm when the acquisition time is limited to 30 min.

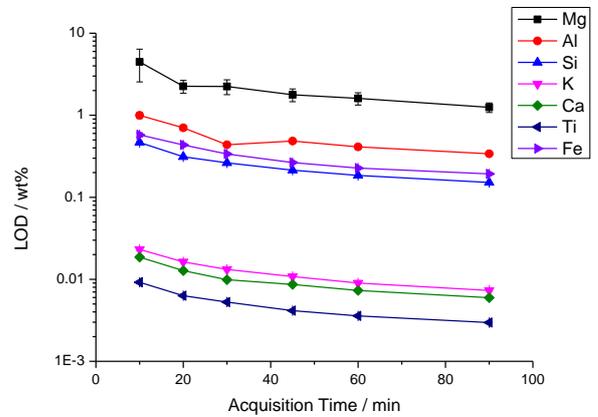

Figure 8. LOD variation with acquisition time

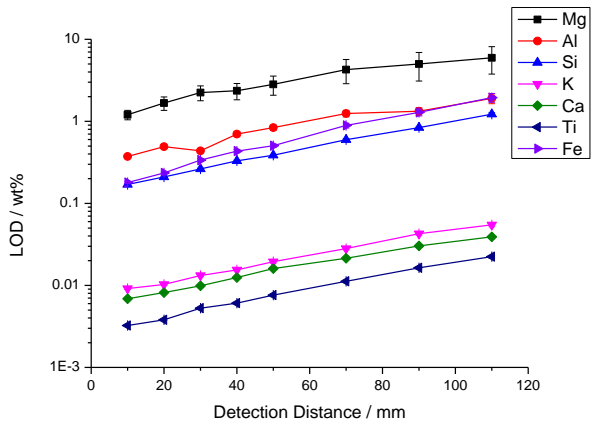

Figure 9. LOD variation with detection distance

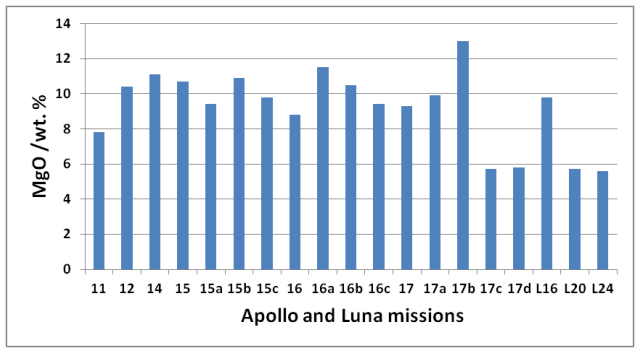

Figure 10. MgO content of average soils at lunar landing sites and in selected regions at the Apollo 15, Apollo 16, Apollo 17, Luna 16, Luna 20, and Luna 24 sites.

## 4 Conclusion

According to the ground-based verification tests, we can come to the following three conclusions.

First, when we set the detection distance to be 30 mm and acquisition time to be 30 minutes, the seven major elements (Mg, Al, Si, K, Ca, Ti, Cr, Fe) can be easily identified in the spectra and the relative deviation of concentration for each major element is less than 15 wt. %.

Second, increasing the acquisition time helps to improve the detection limit for each element. Under the specific verification test conditions (activity of radioactive sources, vacuum degree, temperature and so on), the acquisition time should be longer than 20 min in order to give good detection of Mg element at the distance of 30 mm.

Finally, the detection limits increase with increasing detection distance. Under the specific verification test conditions, if the detection distance is more than 50 mm, the presence of low atomic number elements, such as Mg, in the unknown samples cannot be effectively determined yet.